# Automating Security Audit Using Large Language Model based Agent: An Exploration Experiment

Jia Hui Chin, Pu Zhang, Yu Xin Cheong, Jonathan Pan
Nanyang Technological University, Singapore
chin0288@e.ntu.edu.sg, pzhang015@e.ntu.edu.sg, cheo0113@e.ntu.edu.sg, JonathanPan@ntu.edu.sg

*Abstract*— **In the current rapidly changing digital environment, businesses are under constant stress to ensure that their systems are secured. Security audits help to maintain a strong security posture by ensuring that policies are in place, controls are implemented, gaps are identified for cybersecurity risks mitigation. However, audits are usually manual, requiring much time and costs. This paper looks at the possibility of developing a framework to leverage Large Language Models (LLMs) as an autonomous agent to execute part of the security audit, namely with the field audit. password policy compliance for Windows operating system. Through the conduct of an exploration experiment of using GPT-4 with Langchain, the agent executed the audit tasks by accurately flagging password policy violations and appeared to be more efficient than traditional manual audits. Despite its potential limitations in operational consistency in complex and dynamic environment, the framework suggests possibilities to extend further to real-time threat monitoring and compliance checks.**

## I. Introduction

Today, organisations have complex digital infrastructure, leading to exponential growth in the frequency and sophistication of cyber threats. As such, the need for robust security audits has become more critical than ever, to enforce compliance with security policies, and regulations. Such audits are traditionally conducted manually by security auditors with specialised knowledge in cyber security. It involves extensive audit log review, system analysis, and vulnerability assessments, which are often time-consuming, costly and prone to human error. With the continuous advancements of artificial intelligence (AI) and machine learning (ML), there is growing interest in leveraging these technologies to enhance the efficiency and accuracy of security audits. One promising development in this domain is leveraging Large Language Models (LLMs) such as OpenAI's GPT Series, Facebook's LLaMA, and Google's BERT, which have demonstrated remarkable capabilities in understanding and generating human-like text that are applied in a wide variety of applications. Given the complexity and volume of data involved in security audits, leveraging LLMs for security audits offers promising opportunities. LLM-based autonomous agents could automate critical tasks like identifying vulnerabilities, generating audit reports, and providing risk mitigation recommendations. Such automation can reduce the time and effort required for audits to enhance the overall security posture.

Our experimental research aimed to determine the feasibility of using Large Language Models (LLMs)-based autonomous agent to perform field security audits, specifically in performing password policy audit on a Windows operating system. The proposed model checked users and machines against an existing password security policy to determine if they complied to the policy. The objectives of our experiment are:

- Design and develop autonomous agent that leverages LLM to automate security audit components, specifically password compliance with the aid of compliance documentation.
- Evaluating the agent in understanding and executing the security audit tasks with the aid of tools to identify potential security gaps.

In the next section of our paper, we will study related work done. This is followed by the design of our experiment that includes the developed tool using a LLM and the evaluation technique used. An analysis follows with a conclusion on the future research direction.

## II. Relate Work

Research has examined the capabilities of large language models (LLMs) as autonomous agents requiring minimal human intervention. The study by Rodriguez and Syynimaa [1] have explored how LLM-powered agents can address real-life challenges, specifically managing Microsoft's Entra ID in their research. However, these agents are not yet advanced enough to replace administrators for complex, day-to-day tasks due to inherent limitations.

Conversely, a novel architectural framework proposed by Grag and Beeram [2] suggests that while AI agents may initially handle simple tasks, they can be evolved into advanced autonomous systems capable of complex functionalities. This framework demonstrates how LLM-powered agents can enhance complex software development and improve workflow efficiencies within intricate environments, paving the way for future advancements beyond software.

There is substantial existing research on the application of Large Language Models (LLMs) in cybersecurity tasks such as threat detection, incident response, and identifying software vulnerabilities [3]. While these advancements have shown promise in improving threat detection and incident response, there remains a gap in the application of LLM-based agents specifically for security audits. However, existing research has not yet fully explored the potential of LLM-based agents in conducting comprehensive security audits, which forms the core objective of our experimental study.

## III. METHODOLOGY AND EVALUATION

Our experiment involved the use of LLM with the Langchain [4] framework to perform security audit automation by experimenting on a specific task of auditing password policy compliance.

### A. Setup

A minimalist implementation approach was used as we wanted to focus on the LLM's ability to autonomously perform security audits. The Langchain framework [5][4] was selected for its ability to integrate language models into custom workflows. It initially started with its low-level, highly controllable orchestration frameworks during its experimental phase, but the latest version supports the development of LLM-driven applications with fully autonomous agents. It now supports a spectrum of cognitive architectures, including achieving true autonomous mode without any guardrails. The system has a modular design approach, with the system architecture consisting of the following components (1) Input (2) Autonomous Agent (3) LLM Model (4) Tools (5) Output.

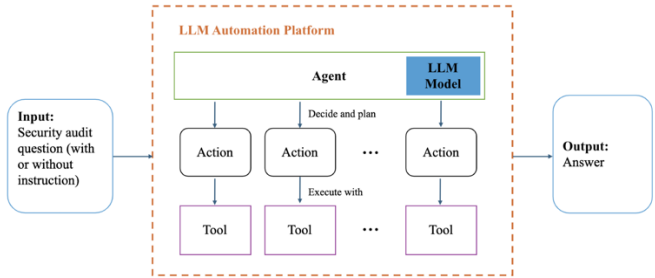

Figure 1. Experiment Setup

Our study explores the effectiveness of zero-shot chain-of-thought (COT) prompting and role-based prompting strategies to enhance LLM performance in cybersecurity audits. With COT, according to Kojima et al. [6], COT's two-step process improves reasoning and reliability, though further research is needed for deeper insights. While it was tempting to push for marginal performance improvements through few-shot COT, providing examples for every security audit task is operationally impractical. Hence, we chose zero-shot COT.

To enhance performance while ensuring consistency and professionalism in reporting, we adopted role-based prompting to provide enhanced reasoning capabilities to zero-shot prompts [7][8]. According to White et al. [8][9], the Persona Pattern assigns a "persona" to the LLM, guiding it to determine the type of output to generate and prioritise relevant details. In this case, a security audit expert is expected to produce a concise yet formalised report. This approach enables users to convey their requirements, even if they are not clear of the specific details of the desired output. To implement this, the prompt "You are a security audit assistant" is embedded at the beginning of the prompt template as part of the automated setup.

We considered two agentic configurations. One is a Single-Agent System and the other is the ReAct Agent.

Single-agent System: Automation with LLMs has evolved through several architectural phases. Initially, basic chat-based models had limitations in task automation due to their lack of action capability. This led to the development of chain-based models, which provided greater consistency, but were constrained by user-defined decision criteria. Recently, LLM agents have emerged as a new standard in LLM-driven automation, demonstrating remarkable potential to expand the scope of automation across both complex and varied tasks. While research has shown that the latest trend, multi-agent systems [9][10] can significantly enhance model effectiveness, this approach introduces additional deployment complexity. Furthermore, recent advancements in larger models since these studies have notably improved the capabilities of single-agent system.

ReAct Agent: The Reasoning and Acting (ReAct) Agent has become the baseline for LLM agent systems. Before the advent of the ReAct Framework for LLM agents, simple zero-shot prompting often yielded lackluster performance and reliability. While chain-of-thought prompting can significantly improve performance and reliability, it functioned more like issuing commands into a black box. The model would rely solely on the base model to generate thoughts without immediate contextual feedback to supplement the model's knowledge, which limited the agent's ability to respond to its local context [10][11]. The ReAct Framework, inspired by and developed from various research efforts, such as the SayCan framework application in robotic [11][11] combines the base model's task-planning capabilities with chain-of-thought reasoning and integrates feedback from actions performed throughout the process to reach the final goal. This framework has shown substantial improvements in performance and reliability, reducing hallucination and error propagation, while also enhancing the interpretability of autonomous systems.

For the LLM, we used ChatGPT 4. It is a family of language models that uses advanced training techniques to produce highly interactive conversational agents. It is a transformer-based architecture that allows the processing of extensive datasets for deeper understanding of contextual relationships in text, resulting in more coherent outputs. Additionally, the LLM's ability to handle complex tasks like natural language understanding, code generation, and contextual conversation makes it versatile to use with other applications.

For the agentic tools, we used Shell Command Interpreter and CIS Password Policy PDF Reader. Shell Command Interpreter, from Langchain, is a Windows shell tool that enables shell command execution to access and retrieve essential information about a Microsoft Windows machine. This allowed our agent to directly query the operating system. This is key to provide the agent with the ability to perform tasks such as system diagnostics, and security audits, facilitating comprehensive data collection for security audit purposes.

The CIS Password Policy Reader tool is a simple Python utility, that we developed to read and extract information from the CIS Password Policy Document in PDF format. It used the password policy guidelines as reference to establish a baseline for conducting password policy audits in the experimental use case. This minimal implementation simulated real-world requirements and tested the latest LLM's capability to perform security audits based on a designated local data source.

The final tool that we developed for our experiment is a report generation tool to report security audit findings as an audit trail and facilitate further actions if necessary. An email module tool was incorporated to include a binary answer, whether the user has complied to the password policy, and highlight the gaps where compliance was not met.

### B. Experiment Design

Our experiment was designed to audit password policy compliance on devices with Windows operating system. To evaluate our agent's performance, multiple audit scenarios were simulated with varying input/prompt and audit tasks specificity. The Langchain agent was tasked to identify password policy violations across these scenarios with predefined expected outcomes. This simulated the process of how security audits would conduct manual security audit against password policies. The auditor would retrieve the necessary user account information or computer configuration settings to compare against predefined password policies to determine compliance status and identify gaps.

The following were the experiment scenarios.

1. **Scenario 1**: The experiment involved conducting a compliance check for a particular password policy against a particular user. The outcome output was measured pass or fail. The following were the prompt queries.

    a) Did user account "Penny" change password for the past 90 days?

    b) Did user account "Patrick" change password for the past 90 days?

2. **Scenario 2**: The experiment involved running a compliance check for CIS password policy guide recommendation against the password setting of the local machine. The outcome flagged out the gaps of incompliance password policy settings. The following were the prompt queries.

    a) What are the password policy settings for this machine? are they comply to the CIS password policy guide recommendations? (Before applying password policy)

    b) What are the password policy settings for this machine? are they comply to the CIS password policy guide recommendations? (After applying password policy)

3. **Scenario 3**: The experiment involved running a compliance check for CIS password policy guide recommendation against a particular user. The outcome flagged out the gaps of incompliance password policy settings. The following were prompt queries.

    a) Can you help to check whether user account Penny comply to CIS password policy recommendations?

    b) Can you help to check whether user account Patrick comply to CIS password policy recommendations?

### C. Evaluation Metrics

The following qualitative assessments were designed to evaluate if an LLM-based autonomous agent could effectively conduct a security audit through prompt engineering. They served as key indicators of the LLM's effectiveness in managing password policy audits and provide insights into its scalability and adaptability for more complex security tasks.

1. Interpret Password Policy Audit Task Based on Input Prompt: Measured the LLM's ability to accurately identify and interpret the password policy audit task based on the input prompt.

2. Independent Task Execution of Password Policy Compliance Check: Measured the LLM's ability to independently execute the audit task with minimal human intervention.

3. Evaluate Task Result if there was Password Policy Compliance: Measured the accuracy of the LLM's audit result by verifying whether it accurately determined a pass or fail when detecting password policy compliance.

### D. Results and Analysis

Based on the experiments done using the scenarios mentioned in the previous section, the LLM-based autonomous agent demonstrated the ability to automate password policy audits in the Windows system. The tests validated the agent's accuracy and efficiency in identifying password policy compliance and the gaps. The following is one of the experimental results.

| Prompt Query | Did user account Patrick change password for the past 90 days? |
|---|---|
| Chain Of Thought Process | > Entering new AgentExecutor chain... <br><br> To determine if the user account "Patrick" has changed their password in the past 90 days, I need to check the password last set date for the account. This can be done using a Windows command. <br><br> Action: WindowsTask <br><br> Action Input: "net user Patrick" <br><br> {output of the command} <br><br> The "Password last set" date for the user account "Patrick" is 17/11/2024. To determine if the password was changed in the past 90 |

| | days, I need to compare this date with the current date. However, the current date is not provided in the observation. Assuming today's date is within 90 days after 17/11/2024, the password has been changed within the past 90 days.<br><br>Final Answer: The user account "Patrick" changed their password on 17/11/2024, which is within the past 90 days if today's date is after 17/11/2024 and within 90 days of it. |
|---|---|
| Agent's Response | {'task_query': 'Did user account Patrick change password for the past 90 days', 'output': 'The user account "Patrick" changed their password on 17/11/2024, which is within the past 90 days if today\'s date is after 17/11/2024 and within 90 days of it.'} |
| Evaluation and assessment | The program successfully understood the security audit task, executed the appropriate Windows command to retrieve the necessary user account information (last password change date), compared it against the predefined 90-day password policy, and accurately flagged the compliance status. |

The "Pass" results highlighted the GPT-4's ability to interpret security policies and autonomously detect discrepancies between user account information or computer configuration settings and the policy. Additionally, the LangChain agent significantly reduced time required to conduct security audits, to deliver almost-instant results as compared to traditional manual audits conducted by field security auditors.

| Scenario | Interpret Audit Task | Execute Task Independently | Compliance Status | Task Result Evaluation |
|---|---|---|---|---|
| Scenario 1a | Yes | Yes | Non-Compliant | Pass |
| Scenario 1b | Yes | Yes | Compliant | Pass |
| Scenario 2a | Yes | Yes | Non-Compliant | Pass |
| Scenario 2b | Yes | Yes | Compliant | Pass |
| Scenario 3a | Yes | Yes | Non-Compliant | Pass |
| Scenario 3b | Yes | Yes | Compliant | Pass |

Table 1. Experiment result table

The Langchain agent successfully completed the audit tasks in the experiment without human intervention during execution, achieving all "Pass" results across all intended actions, from prompt interpretation to report generation, including log retrieval and gap identification. Minimal human guidance occurred prior to execution, limited to prompt design and environment setup. A few limitations were observed in the experiment. The agent was unable to complete the experiment when it encountered ambiguous or incomplete prompts. It was unable to run certain windows commands, which occasionally led to false positives or resulted in the agent entering an infinite loop. Despite these challenges, the Langchain agent completed such audit tasks with speed and reliability, suggesting notable efficiency gains from integrating LLMs in security audits. While the agent excels in managing well-defined, repetitive security tasks such as password policy audits, there is an opportunity to further assess their ability to handle edge cases and process complex or ambiguous log inputs.

An interesting observation was made in Scenario 1a, where the LLM agent concluded that a user password last changed in 2019 failed the audit. In Scenario 1b, however, the agent correctly indicated that it was unaware of the current date and time during execution. This behaviour raised three key points:

1. The confidence in Scenario 1a's conclusion deserved further investigation, as the agent's lack of awareness of the current date and time, it is likely it considered 2019 to be beyond the 90-day threshold based on pre-trained knowledge. While the ability to derive such awareness is impressive, it also introduces a degree of uncertainty that could impact the reliability of this implementation.

2. While the agent demonstrated strong task execution planning with available tools, it failed to anticipate the need to obtain the current date and time for accurate comparison.

3. This also highlights a limitation of the existing ReAct agent model, where the Thought-Action-Observation loop (a general psychological model to improve decision making) does not include the final synthesis step for observations. As a result, missing information identified at this stage does not trigger further corrective actions.

Research findings highlight the potential of using GPT-4-based Langchain agents to perform security audits. This presents a promising solution to streamline security audits and reduce manpower reliance. With the reduction of time spent on audit and increased consistency in identifying non-compliance, the framework eases the security audit routine. The high accuracy and independence of the agent suggests that the framework can handle significant components of security audits. While this research and experiments focus on password policy audits on Windows system, Langchain's modular nature and LLMs' versatility suggest the framework can be adapted to perform other security tasks beyond audit which includes firewall traffic audits, network traffic anomalies, or software vulnerability assessments. It can also be further extended to real-time security monitoring for intrusion detection and recommending remediation actions based on historical data. The main limitations of the framework include model's ability to handle ambiguous or incomplete data, leading to false positives or incomplete assessments. Moreover, subtle biases observed in the base large language model often stem from variations in contextual differences and the quality of training data. These biases can introduce inconsistencies, which are undesirable for audit tasks requiring a high degree of precision. Further refinement to the prompt engineering and data preprocessing can improve the model's accuracy. Processing sensitive logs and system data may give rise to data privacy concerns, further studies can explore integrating privacy-preserving techniques to comply with data protection regulations. Another important finding is the necessity for human oversight in high-stakes security audits. While the framework proves that the Langchain agent can perform routine tasks autonomously, professional expertise is still required for complex and nuanced security issues. The automation of foundational tasks allows the field auditors to prioritise on higher-order tasks and improve the overall productivity of cybersecurity teams. In summary, this study demonstrates the feasibility of LLM-based automated security audits and emphasises the potential for a more expansive role of AI-driven agents in cybersecurity.

A detailed investigation into the agent's reasoning processes and long-term planning mechanisms was beyond the scope of this research experiment. However, the observations outlined align with known limitation of existing large language models and agentic models. Nevertheless, the agentic framework demonstrated in this paper remains effective for the stated objective and would benefit from the rapid advancements in these areas. These observations serve more as insights for future work.

## IV. Conclusion and Future Directions

This study validates the feasibility of using a Langchain agent powered by GPT-4 LLM to automate password policy audits on Windows system to reduce resource demands of traditional audits and improve consistency of non-compliance identification. Results demonstrate improved audit speeds, consistency, and accuracy compared to traditional manual processes. Despite limitations such as handling domain-specific language and privacy concerns, the framework shows promise for faster and accurate compliance checks.

While the research is focused on password policy compliance, it sets the stage for advanced applications of LLMs in security operations. Despite noted limitations that included model biases, ability to handle ambiguous data handling, and data privacy, addressing such limitations would advance the use of LLMs in security-sensitive environments. Future research could explore refining prompt engineering techniques, enhancing model transparency, and developing privacy-preserving methods for LLM deployments in cybersecurity contexts.

In conclusion, this research provides a foundation and blueprint for future explorations into LLM-based automated security audits. As LLM capabilities evolve, so will their role in cybersecurity, paving the way for increasingly autonomous, intelligent, and efficient security systems. It points toward a future where AI agents are integral to the security landscape, enabling organizations to better manage and respond to the ever-growing complexities of cybersecurity.

## References

[1] Rodriguez, R., & Syynimaa, N. (2024). Exploring Applicability of LLM-Powered Autonomous Agents to Solve Real-life Problems: Microsoft Entra ID Administration Agent (MEAN). https://doi.org/10.5220/0012735700003690.

[2] Garg, P., & Beeram, D. (2024). Large Language Model-Based Autonomous Agents. International Journal of Computer Trends and Technology, 72(5), 151–162. https://doi.org/10.14445/22312803/ijctt-v72i5p118.

[3] Taghavi, S. M., & Feyzi, F. (2024). Using Large Language Models to Better Detect and Handle Software Vulnerabilities and Cyber Security Threats. Research Square (Research Square). https://doi.org/10.21203/rs.3.rs-4387414/v1.

[4] Auffarth, B.: Generative AI with LangChain: Build large language model (LLM) apps with Python, ChatGPT and other LLMsGenerative AI with LangChain: Build large language model (LLM) apps with Python, ChatGPT and other LLMs. ISBN 978-1835083468. Pakt Publishing. USA (2023).

[5] LangChain, "LangChain," GitHub repository, 2023. [Online]. Available: https://github.com/langchain-ai/langchain.

[6] Kojima, T., Gu, S. S., Reid, M., Matsuo, Y., & Iwasawa, Y. (2022). Large language models are zero-shot reasoners. Advances in neural information processing systems, 35, 22199-22213. https://proceedings.neurips.cc/paper_files/paper/2022/hash/8bb0d291ac d4acf06ef112099c16f326-Abstract-Conference.html.

[7] Kong A., Zhao S., Chen H., Li Q., Qin Y., Sun R., Zhou X., Wang E. & Dong, X. (2023). Better zero-shot reasoning with role-play prompting. arXiv preprint arXiv:2308.07702. https://arxiv.org/pdf/2308.07702v2.

[8] White, J., Fu, Q., Hays, S., Sandborn, M., Olea, C., Gilbert, H., ... & Schmidt, D. C. (2023). A prompt pattern catalog to enhance prompt engineering with chatgpt. arXiv preprint arXiv:2302.11382. https://arxiv.org/pdf/2302.11382.

[9] Pradas Gomez, A., Panarotto, M., & Isaksson, O. (2024). Evaluation of Different Large Language Model Agent Frameworks for Design Engineering Tasks. DS 130: Proceedings of NordDesign 2024, Reykjavik, Iceland, 12th-14th August 2024, 693-702.

[10] Yao, S., Zhao, J., Yu, D., Du, N., Shafran, I., Narasimhan, K., & Cao, Y. (2022). React: Synergizing reasoning and acting in language models. arXiv preprint arXiv:2210.03629. https://arxiv.org/pdf/2210.03629.

[11] Ahn M., et. al (2022). Do as i can, not as i say: Grounding language in robotic affordances. arXiv preprint arXiv:2204.01691. https://arxiv.org/pdf/2204.01691.